\documentclass{kapproc} 

\usepackage{procps} 

\usepackage[dvips]{graphicx}

\upperandlowercase

 \let\footnote\savefootnote

\normallatexbib 

\def\kms{km~s$^{-1}$}
\def\farcs{\hbox{$.\!\!^{\prime\prime}$}}
\def\ga{\ifmmode\stackrel{>}{_{\sim}}\else$\stackrel{>}{_{\sim}}$\fi} 
\def\la{\ifmmode\stackrel{<}{_{\sim}}\else$\stackrel{<}{_{\sim}}$\fi} 
\def\etal{{et~al.}}
\def\HI{H\,{\sc i}}

\begin{document}

\articletitle
{Pulsar Bow Shocks as Probes of Warm Neutral Gas}

\chaptitlerunninghead{Pulsar Bow Shocks as Probes of Warm Neutral Gas}

\author{Bryan M. Gaensler,\altaffilmark{1} Benjamin Stappers,\altaffilmark{2,3}
Shami Chatterjee,\altaffilmark{4,5} Parviz Ghavamian,\altaffilmark{6}
D.\ Heath Jones,\altaffilmark{7} \& James Cordes\altaffilmark{4}}

\affil{\altaffilmark{1}Harvard-Smithsonian Center for Astrophysics
\ \altaffilmark{2}ASTRON \ \altaffilmark{3}U Amsterdam \
\altaffilmark{4}Cornell University \ \altaffilmark{5}NRAO \
\altaffilmark{6}Rutgers University \ \altaffilmark{7}ANU}

\begin{abstract}
Pulsars have mean space velocities $\ga$500~\kms. The consequent ram
pressure results in tight confinement of the star's energetic wind,
driving a bow shock into the surrounding medium. Pulsar bow shocks have
long been regarded as a curiosity, but new optical and X-ray observations
are both rapidly expanding the sample of such sources, and are offering
new ways to probe the interaction between pulsars and their environments.
Here we discuss some of these new results, and explain how these data
can be used to probe the density and structure of neutral gas in the 
interstellar medium.

\end{abstract}

\section{Introduction}

Pulsars release their rotational kinetic energy via relativistic winds, with ``spin-down
luminosities'' typically in the range $\dot{E} = 10^{32}
- 10^{38}$~ergs~s$^{-1}$. Pulsars also have high space velocities,
typically $V_{PSR} = 100 - 2000$~\kms, meaning that they are almost
always moving supersonically through surrounding gas. We can
therefore conclude that most pulsars drive bow shocks in the ambient 
interstellar medium (ISM).

Pulsar bow shocks are potentially a very powerful probe of interstellar
gas.  The precision of pulsar timing means that we usually have very accurate
measurements of a pulsar's $\dot{E}$, its position, its proper motion,
and its distance (via either parallax or the dispersion of the pulses).
Thus in a pulsar bow shock, the only remaining unknowns are the
density/structure of the ISM, plus the inclination of the pulsar's
velocity vector to the line of sight. Furthermore, pulsars typically have
ages of $10^6-10^9$~years, so that they are usually well-removed from
star-forming regions and are thus a relatively unbiased tracer of the ISM.

\section{Pulsar Bow Shocks: Theory and Observation}

The pulsar/ISM interaction generates two shocks, a forward shock (the
bow shock) and a reverse shock (the pulsar wind termination shock),
separated by a contact discontinuity.  The fundamental size scale of
the system is the distance along the symmetry axis between the pulsar
and the contact discontinuity.  This ``stand-off distance'', $r_w$, is
set by ram-pressure balance: $\dot{E}/\Omega r_w^2 c = \rho V_{PSR}^2$.
Here $\Omega$ is the solid angle of the outflow in the pulsar wind,
and $\rho$ is the ambient density.  The shape of the bow shock surface
has an analytic solution, as shown by [1].

We observe two distinct types of emission from bow shocks. The
forward shock is often seen in H$\alpha$, resulting from
collisional excitation of neutral hydrogen, plus charge exchange with
protons behind the shock. We also see radio/X-ray synchrotron
emission, produced by relativistic particles accelerated
at the termination shock. Only around PSR~B1957+20 (Fig~\ref{fig}, left)
have we as yet seen both H$\alpha$ and synchrotron emission
in the same source [2].

\section{Size and Morphology}

The value of $r_w$ can be directly estimated from an image of a bow shock. If
we assume that $\Omega \approx 4\pi$, and derive $\dot{E}$ and $V_{PSR}$
from pulsar timing and dispersion, we can use pressure balance to
determine $\rho$. This calculation is complicated by the unknown inclination angle and
other effects, but overall, it can be shown that the known bow shocks
are all consistent with ambient densities $n_0 \approx 0.1$~cm$^{-3}$,
as expected for the warm neutral ISM [3].

The morphology of the bow shock powered by PSR~J0437--4715 is well
described by the idealised form derived by [1], allowing one
to derive the inclination angle of the pulsar's velocity vector to the
line of sight [4]. However, the recently
discovered bow shocks around PSRs J2124--358 [5] and B0740--28 [6]
show asymmetries,
kinks and other features in their H$\alpha$ emission (Fig~\ref{fig},
centre \& right).  These deviations
from the analytic solution can be used to infer the presence of a density
gradient and/or relative flow velocity in the ambient ISM [5, 7].

\begin{figure}[ht]
\vspace{4cm}
\vspace{-5mm}
\caption{Pulsar bow shocks. In each case, the white arrow
indicates the pulsar's projected direction
of motion. {\bf Left:} H$\alpha$ (grayscale)
and X-ray (contours) emission around PSR~B1957+20 [2];
{\bf Centre:} H$\alpha$ emission around PSR~J2124--3358 [5];
{\bf Right:}  H$\alpha$ emission around PSR~B0740--28 [6].}
\label{fig}
\end{figure}

\section{Time Variability}

A pulsar moving at 500~\kms\ at a distance of 1~kpc has a proper
motion of 100~mas~year$^{-1}$. The resulting bow shock motion is easily
detectable in a few years with modern optical facilities. For example,
if a pulsar is moving into a region of increasing density, one 
expects the head to narrow and the stand-off distance to shrink. This
is indeed what has been observed in the ``Guitar Nebula'' powered by
PSR~B2224+65, between two epochs separated by 7 years [3].
Such changes suggest
fluctuations at the level of $\Delta n_0
\sim 1-10$~cm$^{-3}$ on scales $\Delta x \sim 0.02$~pc. 
Further such measurements of
density variations may may provide a useful ``missing link''
between the density fluctuations seen in \HI\ at scales $\sim 0.1-200$~pc
(e.g.\ [8]) and the ``tiny scale atomic structure'' 
at scales $\sim$5--100~AU seen toward pulsars and through VLBI (e.g.\
[9]).

\section{Conclusions}

We are now planning a number of further avenues of investigation:
multi-epoch imaging to better characterise the time-variability of
these sources; deep spectroscopy to identify high-velocity H$\alpha$
emission and thus probe the post-shock flow; modelling of variations in
brightness and thickness seen around the shock; searching for UV lines
from bow shocks to probe pulsars embedded in ionised gas; modelling
of X-ray and radio data from shocked pulsar winds; and deeper optical
searches with Magellan to increase the sample beyond the six optical bow shocks
currently known.  It is important to realise that the number of promising
targets to search is rapidly increasing --- the number of known pulsars
has doubled in the last five years, and probably will do so again in
another five years. While we are only just beginning to explore their
potential, all these considerations argue that pulsar bow shocks are
emerging as an exciting new probe of the ISM.

\begin{acknowledgments}
BMG acknowledges support from
NASA through SAO grant GO2-3041X.
\end{acknowledgments}

\section{Discussion}

\noindent {\it Hester:} We have tried to find H$\alpha$ from pulsar bow
shocks in the past. Good luck! \\

\noindent {\it Gaensler:} The work done in the mid-1980s was carried
out with the Palomar $60''$, using a 15-\AA\ filter and $1\farcs2$
pixels. Our new observations are with a
6.5-metre telescope, a 7-\AA\ filter and $0\farcs07$ pixels.  We thus
expect to be $\ga10$ times more sensitive than previous searches. Indeed
one of our recently discovered bow shocks, PSR~B0740--28 [6],
was a Palomar non-detection. \\

\noindent {\it Benjamin:} How many of the pulsars in the current sample
have parallaxes, and can you associate these with neutral clouds? \\

\noindent {\it Gaensler:} Two bow shock systems have parallaxes:
PSR~J0437--4715 and RX~J18576.5--3754. These
are both nearby ($<200$~pc), so it is difficult to separate
any associated \HI\ clouds from local gas.  \\

\noindent {\it Raymond:} The X-ray trail behind PSR~B1957+20 is surprisingly
narrow. Why? \\

\noindent {\it Gaensler:} This was a surprise to us also; simulations
suggest that this region should be reasonably
broad. We see this narrow tail behind other objects also, e.g.\
PSRs~B1757--24 (``the Duck'') and J1747--2958 (``the Mouse''). This may
represent a nozzle effect or be the result of magnetic collimation.  \\

\noindent {\it Slavin:} Doesn't the analytic expression for the bow-shock shape
assume that the gas has radiatively cooled?  \\

\noindent {\it Gaensler:} Yes.
However, simulations 
show that the head of a pulsar bow shock
still has a shape which is a good match to the analytic expression [10, 11]. 
There are
deviations from the analytic solution in the tail region, however. \\

\end{document}